\documentclass[fleqn,usenatbib]{mnras}

\usepackage{newtxtext,newtxmath}
\usepackage[T1]{fontenc}
\usepackage{graphicx}
\usepackage[svgnames]{xcolor}
\usepackage{bm}
\usepackage[math]{cellspace}
\usepackage{epigraph}
\usepackage{orcidlink}

\usepackage{hyperref}

\DeclareRobustCommand{\VAN}[3]{#2}
\let\VANthebibliography\thebibliography
\def\thebibliography{\DeclareRobustCommand{\VAN}[3]{##3}\VANthebibliography}

\def\nht#1{\textcolor{black}{#1}}

\def\nhttwo#1{\textcolor{black}{#1}}

\title[Strong lensing constraints on dark energy]{\nht{Constraints on dark energy from TDCOSMO \& SLACS lenses}}

\author[N. B. Hogg]
{Natalie B. Hogg \orcidlink{https://orcid.org/0000-0001-9346-4477}\thanks{E-mail: \href{mailto:natalie.hogg@lupm.in2p3.fr}{natalie.hogg@lupm.in2p3.fr}}
\\
Laboratoire Univers et Particules de Montpellier, Université de Montpellier, CNRS, Montpellier, France, 34090
}

\date{Accepted 2024 January 8. Received 2023 December 3; in original form 2023 October 20.}

\pubyear{2023}

\begin{document}
\label{firstpage}
\pagerange{\pageref{firstpage}--\pageref{lastpage}}
\maketitle

\begin{abstract}
Problems with the cosmological constant model of dark energy motivate the investigation of alternative scenarios. I make the first measurement of the dark energy equation of state using the hierarchical strong lensing time delay likelihood provided by TDCOSMO. I find that the combination of seven TDCOSMO lenses and 33 SLACS lenses is only able to provide a \nhttwo{weak constraint} on the dark energy equation of state, $w < -1.75$ \nhttwo{at 68\% confidence}, which nevertheless implies the presence of a phantom dark energy component. When the strong lensing time delay data is combined with a collection of cosmic microwave background, baryon acoustic oscillation and Type Ia supernova data, I find that the equation of state \nhttwo{is $w = -1.025\pm 0.029$.}
\end{abstract}


\section{Introduction} \label{sec:introduction}
The better part of three decades has passed since one of the most profound discoveries in modern cosmology was made: that the expansion rate of the Universe is currently accelerating~\citep{Riess1998, Perlmutter1999}. In the standard model of cosmology, this acceleration is attributed to the cosmological constant $\Lambda$ acting as a dark energy, with an equation of state $w= P/\rho = -1$, where $P$ is the pressure and $\rho$ the energy density of the dark energy. However, the true nature of dark energy, and in particular whether its energy density is actually constant in time, remains a subject of debate~\citep{Escamilla2023}.

Various dark energy models have been proposed as alternatives to the cosmological constant, most of which incorporate a dark energy whose density changes over time~\citep{Copeland}. A model-agnostic way of investigating this dynamical kind of dark energy is to use a parameterisation of the equation of state which allows for deviations from $w=-1$. A popular choice is the Chevalier--Polarski--Linder (CPL) parameterisation~\citep{Chevallier:2000qy, Linder:2002et},
\begin{equation}
    w(a) = w_0 + (1-a)\, w_{a}, \label{eq:cpl}
\end{equation}
a first-order Taylor expansion in the scale factor $a$. The parameters $w_0$ and $w_a$ can thus be constrained with data; deviations from $w_0 = -1$ and $w_a = 0$ would indicate that the energy density of dark energy is evolving in time.

Myriad observational data have been used to place constraints on the equation of state of dark energy. For example, the Planck satellite's measurements of the cosmic microwave background (CMB) temperature anisotropies, E-mode polarisation and CMB lensing, along with baryon acoustic oscillations (BAO) measured by the 6dF Galaxy Survey and SDSS~\citep{Beutler2011, Ross:2014qpa, Alam:2016hwk} and the Pantheon catalogue of Type Ia supernovae (SNIa)~\citep{Scolnic:2017caz}, produced a measurement of $w=-1.028 \pm 0.031$~\citep{Aghanim:2018eyx}. A stated goal of  Stage~IV surveys such as Euclid is to measure the dark energy equation of state with percent-level precision~\citep{Amendola:2016saw}.


\nht{Strong lensing time delays are the measurement of the different arrival times of the multiple images of a gravitationally lensed source object~\citep{Refsdal1964b}. This time delay depends on the angular diameter distances between the objects involved and thus directly probes the expansion rate of the Universe, $H_0$. A distance ladder -- a calibration of luminosity distances, typically to SNIa, using parallax measurements and the period--luminosity relation of Cepheid variable stars -- is not needed to obtain cosmological information from strong lensing time delays, making them, on paper, a powerful probe of low-redshift cosmology.}

Whilst there is no reliance on the distance ladder in strong lensing time delay measurements of $H_0$, there can be a fairly significant dependence on how the mass profile of the lens galaxy is modelled~\citep{Sonnenfeld:2017dca}. This problem has been recognised since the studies of the very first strongly lensed quasar for which time delays were measured, Q0957+561~\citep{Walsh1979, Falco1985}.

As an example of the effect of different lens modelling on the measurement of $H_0$, we can compare the H0LiCOW measurement  of $H_0 = 73.3^{+1.7}_{-1.8}$kms$^{-1}$Mpc$^{-1}$~\citep{Wong:2019kwg}, which comes from six strong lensing time delays and is a 2.4\% precision measurement, with the TDCOSMO measurement, $H_0 = 74.5^{+5.6}_{-6.1}$kms$^{-1}$Mpc$^{-1}$~\citep{Birrer:2020tax}, which comes from the original six H0LiCOW lenses plus one new lens and is a 9\% precision measurement.

The reduction in precision is mainly the result of relaxing certain assumptions in the H0LiCOW lens modelling\nht{, namely making a choice of parameterisation for the lens mass profiles which is maximally degenerate with the so-called mass--sheet degeneracy, which I explain further below.} To combat this increase in uncertainty, the TDCOSMO team added 33 additional lenses from the SLACS catalogue in order to provide more information on the mass profiles of the TDCOSMO lenses. This data and its nuisance parameters were combined with the original seven lens dataset in a hierarchical Bayesian manner, leading to a measurement of $H_0 = 67.4^{+4.1}_{-3.2}$kms$^{-1}$Mpc$^{-1}$, which is in statistical agreement with both the H0LiCOW measurement and the measurement from the seven TDCOSMO lenses alone. 

While~\cite{Lewis:2002jt} studied the theoretical effect of dynamical dark energy on strong lensing time delay measurements of $H_0$, and the TDCOSMO $H_0$ value itself has been used to constrain variations of the fine structure constant~\citep{Colaco:2020wbn}, axion--photon couplings~\citep{Buen-Abad:2020zbd} and interacting dark energy models~\citep{Wang:2021kxc}, the full likelihood of~\cite{Birrer:2020tax} has not yet been used to obtain cosmological constraints beyond those on $H_0$ and the matter density parameter $\Omega_{\rm m}$. \nht{The H0LiCOW likelihood of \cite{Wong:2019kwg} was used to obtain constraints on various extensions to $\Lambda$CDM, including dynamical dark energy, with the previously alluded to caveat that the mass--sheet degeneracy was only broken by the assumption of certain analytic mass density profiles for the time delay lenses, a potential pitfall which the TDCOSMO likelihood avoids.}

In this work, I present the first \nht{constraint on} the dark energy equation of state using the full hierarchical TDCOSMO likelihood. Strong lensing time delays are particularly useful for this task as the constraints in the $w-\Omega_{\rm m}$ plane are typically orthogonal to those coming from standard rulers i.e. the CMB and BAO~\citep{Motta:2021hvl}. I begin by reviewing strong lensing time delays and the construction of the hierarchical TDCOSMO likelihood. I explain my analysis method and then present my results and conclusions.

\section{Strong lensing time delays} \label{sec:sl}
\subsection{Theory} \label{subsec:theory}
Gravitational lensing is the phenomenon which arises due to the deflection of light by massive objects. The strong lensing regime may be defined as that in which multiple images of a single source are produced. This typically occurs on super-galactic scales, with lenses being galaxies and sources being distant and bright objects such as quasars.

Light from the distant source is deflected by the lens galaxy. Different light paths have different lengths, leading to measurable delays between the arrival times of images. This strong lensing time delay is given by
\begin{equation}
    t(\bm{\theta}, \bm{\beta}) =  \frac{(1+z_{\rm od})}{c} \frac{D_{\rm od} D_{\rm os}}{D_{\rm ds}} \left[\frac{(\bm{\theta}-\bm{\beta})^2}{2} - \psi(\bm{\theta})\right],
\end{equation}
where $z_{\rm od}$ is the redshift of the deflector; $D_{\rm od}$, $D_{\rm os}$ and $D_{\rm ds}$ are the angular diameter distances between the observer and deflector, observer and source and deflector and source; $\bm{\theta}$ is the observed image position; $\bm{\beta}$ is the unknown source position; and $\psi(\bm{\theta})$ is the lensing potential, which carries the information about the mass density profile of the deflector. In a spatially flat Universe ($\Omega_{k} = 0$), the time delay is inversely proportional to $H_0$ via the angular diameter distances involved,
\begin{equation}
    D(z) = \frac{c}{H_0(1+z)} \int_0^z \frac{\mathrm{d}z'}{E(z')},
\end{equation}
where $E(z) \equiv H(z)/H_0$ is the dimensionless Hubble rate.

The difficulty faced by all strong lensing time delay inference is that under any arbitrary linear re-scaling of the source position $\bm{\beta} \rightarrow \lambda \bm{\beta}$, image positions $\bm{\theta}$ are preserved. The lens model is also accordingly transformed, $\bm{\alpha} \rightarrow \lambda \bm{\alpha} + (1- \lambda) \bm{\theta}$, where $\bm{\alpha}$ is the deflection angle. This is known as the internal mass--sheet transform or degeneracy~\citep{Falco1985, Schneider:2013sxa, Schneider:2013wga}. The only way that such a degeneracy can be broken and a measurement of $H_0$ made is by obtaining direct knowledge either of the absolute source size or of the lensing potential itself. The former \nht{may be possible by measuring the size of quasar accretion disks using microlensing, but this technique comes with its own difficulties}~\citep{Chan2021}, so in order to \nht{reliably} constrain $H_0$ a choice must be made in how to model the deflector mass density profile.

\subsection{The TDCOSMO likelihood} \label{subsec:likelihood}
Whilst the H0LiCOW team used analytic mass profiles for the deflectors to break the mass-sheet degeneracy, the TDCOSMO work used stellar kinematics data to provide information about the lensing potential. This led to the initial reduction in precision of the $H_0$ measurement compared to the H0LiCOW result. The precision was increased again by the addition of further stellar kinematics data from a set of 33 SLACS lenses, specifically selected for their similarity to the original seven TDCOSMO lenses. Note that these additional lenses do not have time delay information. Furthermore, the combination of data was made under the assumption that the TDCOSMO and SLACS lenses were drawn from the same parent population.

The complete likelihood describing this dataset was constructed hierarchically, meaning that a set of hyperparameters were defined which allowed all constraints related to the mass-sheet degeneracy to be inferred on a population level, whilst the lens and light model parameters, $\bm{\xi}_{\rm mass}$ and $\bm{\xi}_{\rm light}$, could be inferred on a lens-by-lens basis. Thus, all remaining uncertainty about the mass-sheet degeneracy is propagated to the level of the $H_0$ inference.

Following~\cite{Birrer:2020tax}, the posterior distribution of the cosmological parameters of interest, $\bm{\pi}$, given $N$ sets of individual lens data $\mathcal{D}_{i}$ and the model parameters $\bm{\xi}$, is given by
\begin{align}
    P(\bm{\pi} | \{\mathcal{D}_{i}\}_N) 
    &\propto \mathcal{L}(\{\mathcal{D}_{i}\}_N | \bm{\pi}) P(\bm{\pi}), \nonumber \\
    &= \int \mathcal{L}(\{\mathcal{D}_{i}\}_N | \bm{\pi}, \bm{\xi}) P(\bm{\pi}, \bm{\xi}), \nonumber \\
    &= \int \prod_{i}^N \mathcal{L}(\mathcal{D}_{i}| \bm{\pi}, \bm{\xi}) P(\bm{\pi}, \bm{\xi}).
\end{align}

The nuisance parameter $\bm{\xi}$ is divided into the mass and light model parameters which are constrained at the level of each individual lens, and the set of mass-sheet degeneracy hyperparameters which are constrained at the population level, $\bm{\xi}_{\rm pop}$. The hierarchical TDCOSMO likelihood is thus given by
\begin{align}
    \mathcal{L}(\mathcal{D}_{i}| D, \bm{\xi}_{\rm pop}) = \int &\mathcal{L}(\mathcal{D}_{i}| D, \bm{\xi}_{\rm pop}, \bm{\xi}_{\rm mass}, \bm{\xi}_{\rm light}) \nonumber \\
    &\times P(\bm{\xi}_{\rm mass}, \bm{\xi}_{\rm light})\; \mathrm{d} \bm{\xi}_{\rm mass} \, \mathrm{d} \bm{\xi}_{\rm light},
    \label{eq:likelihood}
\end{align}
where $D$ is the set of angular diameter distances $\{D_{\rm od}, D_{\rm os}, D_{\rm ds}\}$ from which all cosmological results are obtained. For a complete discussion of the construction of the hierarchical likelihood, including the details of the population hyperparameters, I refer the reader to Section 3 of \cite{Birrer:2020tax}.

\nhttwo{The TDCOSMO measurement of $H_0 = 67.4^{+4.1}_{-3.2}$kms$^{-1}$Mpc$^{-1}$} is thus the most precise measurement possible from strong lensing time delays which does not make any assumptions about the deflector mass profiles in order to artificially break the mass-sheet degeneracy. It is also the first strong lensing time delay measurement of $H_0$ which used information from other datasets to improve the precision of the measurement. I will now discuss how I used this hierarchical likelihood to measure the dark energy equation of state.

\section{Method} \label{sec:method}
I wrote an external likelihood package for the cosmological modelling and sampling software \texttt{Cobaya}~\citep{Torrado:2020dgo}, so that the hierarchical TDCOSMO likelihood can be used to obtain constraints on cosmological model parameters in combination with any other cosmological likelihood and with any choice of Boltzmann code. This package is publicly available.\footnote{\url{https://github.com/nataliehogg/tdcosmo_ext}.}

Using the Markov chain Monte Carlo sampler provided by \texttt{Cobaya}, which is adapted from \texttt{CosmoMC}~\citep{Lewis:2002ah, Lewis:2013hha} and uses a fast-dragging procedure to increase sampling speed~\citep{Neal:2005}, I obtained constraints on three cosmological models: $\Lambda$CDM, $w$CDM and $w_0 w_a$CDM. The $w$CDM model allows for a dark energy with a constant equation of state that may differ from $w=-1$; the $w_0 w_a$CDM model allows for a dynamical dark energy. 
In a $w$CDM cosmology (extendable to $w_0 w_a$CDM by the CPL parameterisation shown in Equation \ref{eq:cpl}), and recalling that I keep $\Omega_{k}=0$, the dimensionless Hubble rate is given by
\begin{equation}
    E(z) = \left[\Omega_{\rm m}(1+z)^3 + \Omega_{\rm DE}(1+z)^{3(1+w)}\right]^{\frac12},
\end{equation}
where $\Omega_{\rm m} = \Omega_{\rm b} + \Omega_{\rm c}$, the sum of the  baryon and cold dark matter densities, and $\Omega_{\rm DE}$ is the dimensionless energy density of dark energy today.

For each cosmological model I considered, I sampled the posterior distributions of $H_0$, $\Omega_{\rm b}h^2$ and $\Omega_{\rm c}h^2$, along with the relevant dark energy equation of state parameter(s). In each case, I also sampled the posterior distributions of the hyperparameters associated with the likelihood, and marginalised over them to obtain the posterior distributions on the cosmological parameters of interest. The computation of the angular diameter distances required by the TDCOSMO likelihood was done using the Boltzmann code \texttt{CAMB}~\citep{Lewis:1999bs, Howlett:2012mh}, but I emphasise that I designed the \texttt{Cobaya} interface so that any theory code currently available in or added to \texttt{Cobaya} in the future can be used for this task. Lastly, I used the Parameterised Post-Friedmann framework in \texttt{CAMB} to allow $w$ to cross $-1$~\citep{Fang:2008sn}. 

I validated my implementation of the likelihood package in \texttt{Cobaya} by comparing my $\Lambda$CDM results with the original TDCOSMO results, finding an excellent agreement in terms of constraints on both the cosmological and the nuisance parameters. This validation test, plus all of the code needed to reproduce the results and figures in this paper is also publicly available.\footnote{\url{https://github.com/nataliehogg/slide}.}

Besides the TDCOSMO and SLACS datasets, I also obtained constraints on the models with a dataset consisting of the Planck 2018 measurements of the CMB temperature, polarisation and lensing~\citep{Aghanim:2019ame, Aghanim:2018oex}; the BAO measurements from the 6dF Galaxy Survey~\citep{Beutler2011}, the SDSS Main Galaxy Sample~\citep{Ross:2014qpa} and the SDSS DR12 consensus catalogue~\citep{Alam:2016hwk}; the Pantheon catalogue of Type Ia supernovae \citep{Scolnic:2017caz}; and the TDCOSMO + SLACS lenses, in order to compare with the constraints obtained just using the strong lensing time delay data and with those of the Planck collaboration~\citep{Aghanim:2018eyx}. I will refer to this combination of data as the ``full combination'' from now on.

The priors I used for the cosmological parameters are listed in Table \ref{tab:priors}. Following~\cite{Birrer:2020tax}, I also used the Pantheon prior on $\Omega_{\rm m} = \mathcal{N}(0.298, 0.022)$ when using the TDCOSMO and SLACS data alone, though it is important to note that in my analysis $\Omega_{\rm m}$ was not directly sampled since it is treated as a derived parameter in \texttt{CAMB}, the posterior being obtained from those of $\Omega_{\rm b}$ and $\Omega_{\rm c}$. In the dynamical dark energy cases, I ensured that acceleration occurs by setting a prior on the dark energy equation of state such that $w < -\frac13$. 

\begin{table}
\centering
\begin{tabular}{SlSl}
	\hline
	\hline
	Parameter                    & Prior \\
	\hline
	$\Omega_bh^2$                &  $[0.005,0.1]$\\
	$\Omega_ch^2$                &  $[0.001,0.99]$ \\
	$H_0$~\nhttwo{[kms$^{-1}$Mpc$^{-1}$]} &  $[0, 150]$\\
	$w$ and $w_0$                &  $\left[-3.0, -\frac13\right]$\\
    $w_a$                        &  $[-4.0, 4.0]$ \\
	\hline
\end{tabular}
\caption{Prior ranges of the parameters sampled in my analysis.}\label{tab:priors}
\end{table}

\section{Results} \label{sec:results}
In this section, I present the constraints obtained on $\Lambda$CDM, $w$CDM and $w_0w_a$CDM using the hierarchical TDCOSMO likelihood and the full combination of data listed above. Given the smaller uncertainty on $H_0$ that comes from combining the TDCOSMO + SLACS lenses, I will only present the TDCOSMO alone results for $\Lambda$CDM, to demonstrate the replication of~\cite{Birrer:2020tax}. For the rest of the results, I will show constraints obtained using the complete TDCOSMO + SLACS dataset of 40 lenses. I used \texttt{GetDist} to make the figures and to compute the marginalised parameter values quoted in the text~\citep{Lewis:2019xzd}. \nht{The dark shaded regions of the contour plots represent the 68\% confidence limit and the light shaded regions the 95\% confidence limit.}

\subsection{$\Lambda$CDM} \label{subsec:lcdm}
\begin{figure}
    \centering
    \includegraphics{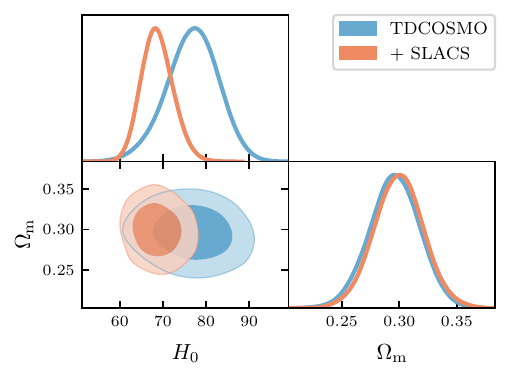}
    \caption{The one and two-dimensional marginalised posterior distributions of $H_0$~\nhttwo{[kms$^{-1}$Mpc$^{-1}$]} and $\Omega_{\rm m}$ in a $\Lambda$CDM cosmology from the seven TDCOSMO lenses (blue) and the \nhttwo{seven TDCOSMO + 33 SLACS}  lenses (red).}
    \label{fig:lcdm}
\end{figure}

In Figure \ref{fig:lcdm}, I show the one and two-dimensional marginalised posterior distributions of $H_0$ and $\Omega_{\rm m}$ in a $\Lambda$CDM cosmology. The blue contours show the constraints from the seven TDCOSMO lenses alone, whilst the red contours show the constraints from the full dataset of 40 lenses (\nhttwo{seven TDCOSMO + 33 SLACS lenses}). As expected, this result replicates the findings of the TDCOSMO paper, with $H_0 = 76.8^{+6.4}_{-5.6}$kms$^{-1}$Mpc$^{-1}$ from the TDCOSMO lenses and $H_0 = 68.7^{+3.4}_{-3.9}$ from the TDCOSMO + SLACS lenses. These values are fully consistent at $1\sigma$ with the TDCOSMO result.

\subsection{$w$CDM} \label{subsec:wcdm}
\begin{figure}
    \centering
    \includegraphics{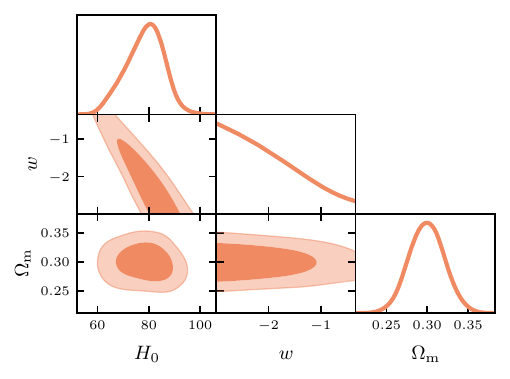}
    \caption{The one and two-dimensional marginalised posterior distributions of $H_0$~\nhttwo{[kms$^{-1}$Mpc$^{-1}$]}, $w$ and $\Omega_{\rm m}$ in a $w$CDM cosmology from the \nhttwo{seven TDCOSMO + 33 SLACS}  lenses.}
    \label{fig:wcdm}
\end{figure}

In Figure \ref{fig:wcdm}, I show the one and two-dimensional marginalised  posterior distributions of $H_0$, $w$ and $\Omega_{\rm m}$ in a $w$CDM cosmology from the \nhttwo{seven TDCOSMO + 33 SLACS}  lenses. The likelihood is able to provide \nhttwo{a weak constraint} on the equation of state of dark energy, $w < -1.75$ \nht{at 68\% confidence}, which implies that the dark energy is `phantom', the term applied when $w <-1$. 

Some consideration must be given here to what is, at first glance, a surprising result. It is known that in some datasets, such as Planck 2018, large negative values of $w$ correlate with large positive values of $H_0$~\citep{Escamilla2023}. This is due to the so-called geometrical degeneracy, where $H_0$, $\Omega_{\rm m}$ and $w$ can take various values which in combination lead to the same value for the angular diameter distance to the surface of last scattering and hence the same angular size of the sound horizon, provided the physical sound horizon size is kept fixed~\citep{Efstathiou:1998xx}.

Since strong lensing time delays also rely on angular diameter distances to probe cosmology, I infer that a similar degeneracy exists here. This conclusion is supported by the clear correlation between $H_0$ and $w$ in Figure \ref{fig:wcdm}. Thus, the strongly negative dark energy equation of state is likely driven by the high central value of $H_0 = 78.4^{+8.3}_{-6.3}$ obtained in this case -- which is nevertheless consistent with the $\Lambda$CDM value at $1\sigma$. \nht{Furthermore, the 95\% confidence limit is $w < -0.74$, reflecting the broadness of the constraint obtained.} As expected, based on the results of~\cite{Birrer:2020tax}, the marginalised posterior value of the matter density $\Omega_{\rm m}=0.299\pm 0.022$ is largely informed by the Pantheon prior. This also acts to somewhat ameliorate the aforementioned degeneracy. 

Lastly, I note that a deeply phantom equation of state (for this value of $\Omega_{\rm m}$, $w\lesssim-1.4$) corresponds to a violation of the null energy condition~\citep{Colgain:2021beg}, which may be problematic depending on the specific cosmological model~\citep{Rubakov:2014jja}.

\subsection{$w_0 w_a$CDM} \label{subsec:w0wacdm}
\begin{figure}
    \centering
    \includegraphics{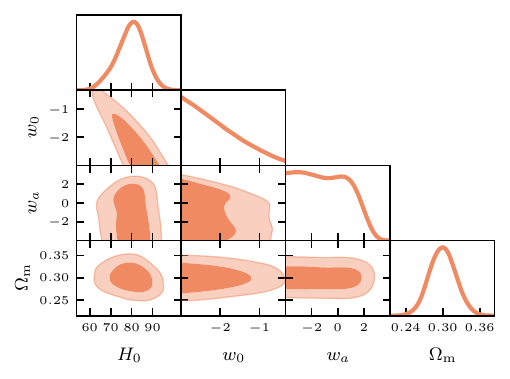}
    \caption{The one and two-dimensional marginalised posterior distributions of $H_0$~\nhttwo{[kms$^{-1}$Mpc$^{-1}$]}, $w_0$, $w_a$ and $\Omega_{\rm m}$ in a $w_0 w_a$CDM cosmology from the \nhttwo{seven TDCOSMO + 33 SLACS}  lenses.}
    \label{fig:w0wa}
\end{figure}

In Figure \ref{fig:w0wa}, I show the one and two-dimensional marginalised  posterior distributions of $H_0$, $w_0$, $w_a$ and $\Omega_{\rm m}$ in a $w_0w_a$CDM cosmology from the \nhttwo{seven TDCOSMO + 33 SLACS}  lenses. The likelihood is again only able to provide \nhttwo{loose} upper \nht{limits} on the dark energy equation of state parameters, $w_0 < -1.86$ and $w_a < 0.102$ \nht{at 68\% confidence, and $w_0 < -0.861$, $w_a < 1.97$ at 95\% confidence}. Again the value of $H_0$ is larger than but still consistent with the $\Lambda$CDM measurement: $H_0 = 79.6^{+7.5}_{-6.0}$. 

\begin{figure}
    \centering
    \includegraphics{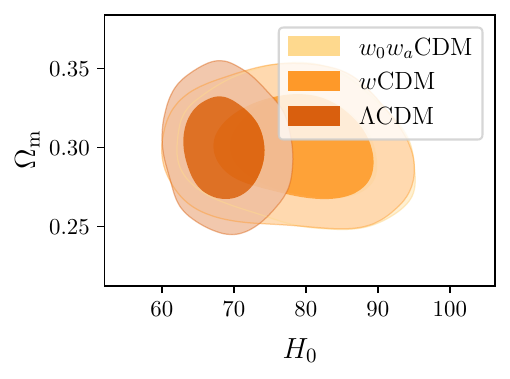}
    \caption{The two-dimensional marginalised posterior distributions of $H_0$~\nhttwo{[kms$^{-1}$Mpc$^{-1}$]} and $\Omega_{\rm m}$ in the three cosmologies studied from the \nhttwo{seven TDCOSMO + 33 SLACS} lenses.}
    \label{fig:H0}
\end{figure}

In Figure \ref{fig:H0}, I show the two-dimensional marginalised  posterior distributions of $H_0$ and $\Omega_{\rm m}$ in a $\Lambda$CDM cosmology (red), a $w$CDM cosmology (orange) and a $w_0w_a$CDM cosmology (yellow) from the \nhttwo{seven TDCOSMO + 33 SLACS}  lenses. From this plot, it is clear that whilst the $H_0$ values in the extended cosmologies are large, they are still consistent at $1\sigma$ with the $\Lambda$CDM result.

\nhttwo{It is important to note that my results are very similar to those found by \cite{Wong:2019kwg}. This similarity implies that, for the six original H0LiCOW lenses, the true lensing potential of each lens is well-approximated by the analytic profiles used in that work i.e. the mass--sheet degeneracy is making little contribution to the uncertainty on the cosmological parameters. Nevertheless, this may not be true for every lens in the Universe, and therefore the hierarchical likelihood procedure developed in \cite{Birrer:2020tax} is the one which should be used for future cosmological inference involving strong lensing time delays.}

\subsection{Full combination of data} \label{subsec:full}
\begin{figure}
    \centering
    \includegraphics{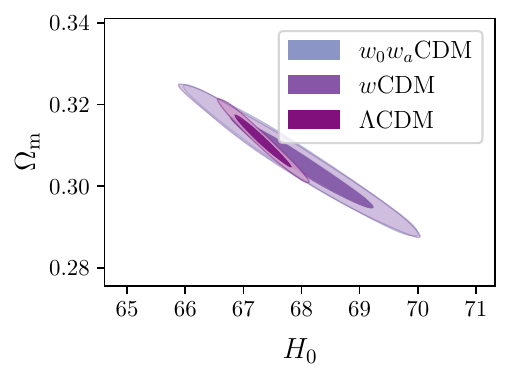}
    \caption{The two-dimensional marginalised posterior distributions of $H_0$~\nhttwo{[kms$^{-1}$Mpc$^{-1}$]} and $\Omega_{\rm m}$ in the three cosmologies studied from the full combination of Planck 2018 + BAO + Pantheon SNIa + TDCOSMO + SLACS.}
    \label{fig:H0_full}
\end{figure}

In Figure \ref{fig:H0_full}, I show the two-dimensional marginalised  posterior distributions of $H_0$ and $\Omega_{\rm m}$ in a $\Lambda$CDM cosmology (dark purple), a $w$CDM cosmology (medium purple) and a $w_0w_a$CDM cosmology (light purple) from the full combination of Planck 2018 + BAO + Pantheon SNIa + TDCOSMO + SLACS data. In this case, I did not use the Pantheon prior on $\Omega_{\rm m}$, since the inclusion of the Pantheon dataset provides the same information as that prior. From this plot we can see that the values of $H_0$ and $\Omega_{\rm m}$ in the extended cosmologies are completely consistent with the $\Lambda$CDM values at $1\sigma$ when using the full combination of data; the $w$CDM and $w_0w_a$CDM constraints are virtually identical.

Furthermore, this combination of data inevitably provides a much stronger constraint on $w$, $w_0$ and $w_a$ than the TDCOSMO + SLACS data alone, and removes any hint of a phantom dark energy component. In  a $w$CDM cosmology, the dark energy equation of state is measured to be $w = -1.025\pm 0.029$, a marginal increase in precision compared to the Planck 2018 + BAO + SNIa value of $w=-1.028 \pm 0.031$. Both of these measurements are consistent with a cosmological constant.  

\begin{figure}
    \centering
    \includegraphics{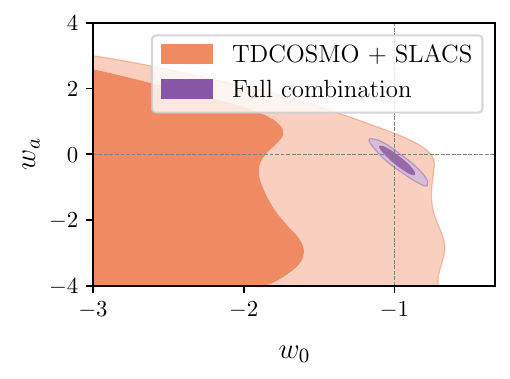}
    \caption{The two-dimensional marginalised posterior distributions of $w_0$ and $w_a$ from the TDCOSMO + SLACS data (red) and from the full combination of Planck 2018 + BAO + Pantheon SNIa + TDCOSMO + SLACS  (purple). The dashed lines show the values of these parameters which correspond to a cosmological constant, $w_0=-1$, $w_a=0$.}
    \label{fig:w0wa_2d}
\end{figure}

In a $w_0w_a$CDM cosmology, $w_0 =-0.985^{+0.071}_{-0.091} $ and $w_a= -0.18^{+0.33}_{-0.25}$, compared to $w_0 =-0.957\pm 0.08 $ and $w_a= -0.29 ^{+0.32}_{-0.26}$ from Planck 2018 + BAO + SNIa. These are again both almost equivalent in precision and consistent with a cosmological constant. This can be clearly seen in Figure \ref{fig:w0wa_2d}, where I show the two-dimensional marginalised posterior distributions of $w_0$ and $w_a$ from the TDCOSMO + SLACS data (red) and from the full combination of Planck 2018 + BAO + Pantheon SNIa + TDCOSMO + SLACS  (purple). The dashed lines in this plot show the values of the dark energy equation of state parameters which correspond to a cosmological constant, $w_0=-1$ and $w_a = 0$.

\begin{table}
\centering
\begin{tabular}{Slllll}
	\hline
	\hline
    Model        & Data             & $\chi^2$ &d.o.f. & $\Delta \chi^2$  \\
	\hline
    $\Lambda$CDM & TDCOSMO + SLACS  & 114.2 &9  &~~ ---\\
    $w$CDM       & TDCOSMO + SLACS  & 113.7 &10 &$-0.5$   \\
    $w_0w_a$CDM  & TDCOSMO + SLACS  & 113.5 &11 &$-0.7$ \\
    $\Lambda$CDM & Full combination &4011.0 &30 &~~ ---\\
	  $w$CDM       & Full combination &4011.0 &31 &~~ $0.0$\\
    $w_0w_a$CDM  & Full combination &4009.0 &32 &$-2.0$ \\
	\hline
\end{tabular}
\caption{The $\chi^2$, degrees of freedom, and $\Delta \chi^2$ values for each case studied.}\label{tab:chi2}
\end{table}

Lastly, I computed the $\Delta \chi^2 \equiv \chi^2_{\rm} - \chi^2_{\Lambda \rm{CDM}} $ for each case studied. The computed values are shown in Table \ref{tab:chi2}. Whilst the $\Delta \chi^2$ is negative for \nht{both extended cosmologies in the TDCOSMO + SLACS cases}, implying that they are favoured over $\Lambda$CDM, the significance of this decrease in $\chi^2$ must be evaluated using a difference table due to the greater number of degrees of freedom in the $w$CDM cosmology with respect to $\Lambda$CDM. With one additional degree of freedom, and for a 95\% level of significance, a $|\Delta \chi^2| > 3.841$ is required for the improvement in fit to be considered significant. This is clearly not the case here. From these results it is also evident that $\Lambda$CDM is a better fit to the \nht{full combination of} data than the \nht{extended} cosmologies, since the $\Delta \chi^2$ is on parity with $\Lambda$CDM \nht{in the $w$CDM case} and still does not exceed $3.841$ for \nht{the $w_0w_a$CDM case}.

\section{Conclusions} \label{sec:conclusions}
In this work, I presented the first constraints on the equation of state of dark energy from the seven TDCOSMO lenses plus 33 SLACS lenses using the hierarchical likelihood provided by TDCOSMO. I wrote an external likelihood package for the \texttt{Cobaya} software, making this likelihood readily available for public use in combination with other cosmological likelihoods and theory codes. 

I replicated the original TDCOSMO results in a $\Lambda$CDM cosmology and then explored two extended cosmologies, finding that the TDCOSMO + SLACS data was not able to place strong constraints on the equation of state of dark energy, obtaining only \nhttwo{weak} upper \nht{limits} on $w$, $w_0$ and $w_a$. The \nht{constraints} I obtained all implied the presence of a phantom dark energy component, $w<-1$, \nht{at 68\% confidence}.

The use of the TDCOSMO likelihood in combination with the Planck 2018 likelihood, BAO data and the Pantheon SNIa catalogue yielded more precise constraints, with all the dark energy equation of state parameters consistent with a cosmological constant, $w=-1$. I computed the $\Delta \chi^2$ to evaluate the fit of each model, finding no preference for the extended cosmologies over the $\Lambda$CDM case. 


In conclusion, while strong lensing time delays are beginning to provide a competitive (albeit lens-modelling-dependent) measurement of $H_0$, it is clear \nht{from the results of this work, which, in line with expectations from previous literature, indicate that other probes are more useful when studying dark energy. To improve the strong lensing time delay constraint, a larger dataset is certainly needed}; perhaps on the order of hundreds or thousands of lenses~\citep{Shiralilou:2019div}. Furthermore, an increase in precision on the $H_0$ inference will naturally lead to a reduction in the effect of the geometrical degeneracy which is contributing to the hints of phantom dark energy in the TDCOSMO data. Fortunately a number of current or near-future experiments, such as JWST, Roman, LSST and Euclid are likely to provide such data in abundance~\citep{Collett:2015roa}.

\section*{Acknowledgements}
I am grateful to Simon~Birrer, Martin~Millon and Judit~Prat for valuable discussions about the TDCOSMO likelihood, and to Pierre~Fleury for his comments on the manuscript.

\section*{Data Availability}

All data associated with this article is publicly available.

\bibliographystyle{mnras}
\bibliography{slide} 

\begin{thebibliography}{}
\makeatletter
\relax
\def\mn@urlcharsother{\let\do\@makeother \do\$\do\&\do\#\do\^\do\_\do\%\do\~}
\def\mn@doi{\begingroup\mn@urlcharsother \@ifnextchar [ {\mn@doi@} {\mn@doi@[]}}
\def\mn@doi@[#1]#2{\def\@tempa{#1}\ifx\@tempa\@empty \href {http://dx.doi.org/#2} {doi:#2}\else \href {http://dx.doi.org/#2} {#1}\fi \endgroup}
\def\mn@eprint#1#2{\mn@eprint@#1:#2::\@nil}
\def\mn@eprint@arXiv#1{\href {http://arxiv.org/abs/#1} {{\tt arXiv:#1}}}
\def\mn@eprint@dblp#1{\href {http://dblp.uni-trier.de/rec/bibtex/#1.xml} {dblp:#1}}
\def\mn@eprint@#1:#2:#3:#4\@nil{\def\@tempa {#1}\def\@tempb {#2}\def\@tempc {#3}\ifx \@tempc \@empty \let \@tempc \@tempb \let \@tempb \@tempa \fi \ifx \@tempb \@empty \def\@tempb {arXiv}\fi \@ifundefined {mn@eprint@\@tempb}{\@tempb:\@tempc}{\expandafter \expandafter \csname mn@eprint@\@tempb\endcsname \expandafter{\@tempc}}}

\bibitem[\protect\citeauthoryear{Aghanim et~al.}{Aghanim et~al.}{2020a}]{Aghanim:2019ame}
Aghanim N.,  et~al., 2020a, \mn@doi [Astronomy \& Astrophysics] {10.1051/0004-6361/201936386}, 641, A5

\bibitem[\protect\citeauthoryear{Aghanim et~al.}{Aghanim et~al.}{2020b}]{Aghanim:2018eyx}
Aghanim N.,  et~al., 2020b, \mn@doi [Astronomy \& Astrophysics] {10.1051/0004-6361/201833910}, 641, A6

\bibitem[\protect\citeauthoryear{Aghanim et~al.}{Aghanim et~al.}{2020c}]{Aghanim:2018oex}
Aghanim N.,  et~al., 2020c, \mn@doi [Astronomy \& Astrophysics] {10.1051/0004-6361/201833886}, 641, A8

\bibitem[\protect\citeauthoryear{Alam et~al.}{Alam et~al.}{2017}]{Alam:2016hwk}
Alam S.,  et~al., 2017, \mn@doi [Monthly Notices of the Royal Astronomical Society] {10.1093/mnras/stx721}, 470, 2617

\bibitem[\protect\citeauthoryear{Amendola et~al.}{Amendola et~al.}{2018}]{Amendola:2016saw}
Amendola L.,  et~al., 2018, \mn@doi [Living Reviews in Relativity] {10.1007/s41114-017-0010-3}, 21, 2

\bibitem[\protect\citeauthoryear{{Beutler} et~al.}{{Beutler} et~al.}{2011}]{Beutler2011}
{Beutler} F.,  et~al., 2011, \mn@doi [Monthly Notices of the Royal Astronomical Society] {10.1111/j.1365-2966.2011.19250.x}, \href {http://adsabs.harvard.edu/abs/2011MNRAS.416.3017B} {416, 3017}

\bibitem[\protect\citeauthoryear{Birrer et~al.}{Birrer et~al.}{2020}]{Birrer:2020tax}
Birrer S.,  et~al., 2020, \mn@doi [Astronomy \& Astrophysics] {10.1051/0004-6361/202038861}, 643, A165

\bibitem[\protect\citeauthoryear{Buen-Abad, Fan  \& Sun}{Buen-Abad et~al.}{2022}]{Buen-Abad:2020zbd}
Buen-Abad M.~A.,  Fan J.,   Sun C.,  2022, \mn@doi [Journal of High Energy Physics] {10.1007/JHEP02(2022)103}, 02, 103

\bibitem[\protect\citeauthoryear{{Chan}, {Rojas}, {Millon}, {Courbin}, {Bonvin}  \& {Jauffret}}{{Chan} et~al.}{2021}]{Chan2021}
{Chan} J.~H.~H.,  {Rojas} K.,  {Millon} M.,  {Courbin} F.,  {Bonvin} V.,   {Jauffret} G.,  2021, \mn@doi [Astronomy \& Astrophysics] {10.1051/0004-6361/202038971}, \href {https://ui.adsabs.harvard.edu/abs/2021A&A...647A.115C} {647, A115}

\bibitem[\protect\citeauthoryear{Chevallier \& Polarski}{Chevallier \& Polarski}{2001}]{Chevallier:2000qy}
Chevallier M.,  Polarski D.,  2001, \mn@doi [International Journal of Modern Physics D] {10.1142/S0218271801000822}, 10, 213

\bibitem[\protect\citeauthoryear{Cola\c{c}o, Gonzalez  \& Holanda}{Cola\c{c}o et~al.}{2021}]{Colaco:2020wbn}
Cola\c{c}o L.~R.,  Gonzalez J.~E.,   Holanda R. F.~L.,  2021, \mn@doi [The European Physical Journal C] {10.1140/epjc/s10052-021-09319-x}, 81, 533

\bibitem[\protect\citeauthoryear{Colg\'ain \& Sheikh-Jabbari}{Colg\'ain \& Sheikh-Jabbari}{2021}]{Colgain:2021beg}
Colg\'ain E.~O.,  Sheikh-Jabbari M.~M.,  2021, \mn@doi [Classical and Quantum Gravity] {10.1088/1361-6382/ac1504}, 38, 177001

\bibitem[\protect\citeauthoryear{Collett}{Collett}{2015}]{Collett:2015roa}
Collett T.~E.,  2015, \mn@doi [The Astrophysical Journal] {10.1088/0004-637X/811/1/20}, 811, 20

\bibitem[\protect\citeauthoryear{{Copeland}, {Sami}  \& {Tsujikawa}}{{Copeland} et~al.}{2006}]{Copeland}
{Copeland} E.~J.,  {Sami} M.,   {Tsujikawa} S.,  2006, \mn@doi [International Journal of Modern Physics D] {10.1142/S021827180600942X}, 15, 1753

\bibitem[\protect\citeauthoryear{Efstathiou \& Bond}{Efstathiou \& Bond}{1999}]{Efstathiou:1998xx}
Efstathiou G.,  Bond J.~R.,  1999, \mn@doi [Monthly Notices of the Royal Astronomical Society] {10.1046/j.1365-8711.1999.02274.x}, 304, 75

\bibitem[\protect\citeauthoryear{{Escamilla}, {Giar{\`e}}, {Di Valentino}, {Nunes}  \& {Vagnozzi}}{{Escamilla} et~al.}{2023}]{Escamilla2023}
{Escamilla} L.~A.,  {Giar{\`e}} W.,  {Di Valentino} E.,  {Nunes} R.~C.,   {Vagnozzi} S.,  2023 (\mn@eprint {arXiv} {2307.14802})

\bibitem[\protect\citeauthoryear{{Falco}, {Gorenstein}  \& {Shapiro}}{{Falco} et~al.}{1985}]{Falco1985}
{Falco} E.~E.,  {Gorenstein} M.~V.,   {Shapiro} I.~I.,  1985, \mn@doi [The Astrophysical Journal Letters] {10.1086/184422}, \href {https://ui.adsabs.harvard.edu/abs/1985ApJ...289L...1F} {289, L1}

\bibitem[\protect\citeauthoryear{Fang, Hu  \& Lewis}{Fang et~al.}{2008}]{Fang:2008sn}
Fang W.,  Hu W.,   Lewis A.,  2008, \mn@doi [Physical Review D] {10.1103/PhysRevD.78.087303}, 78, 087303

\bibitem[\protect\citeauthoryear{Howlett, Lewis, Hall  \& Challinor}{Howlett et~al.}{2012}]{Howlett:2012mh}
Howlett C.,  Lewis A.,  Hall A.,   Challinor A.,  2012, \mn@doi [Journal of Cosmology and Astroparticle Physics] {10.1088/1475-7516/2012/04/027}, 1204, 027

\bibitem[\protect\citeauthoryear{Lewis}{Lewis}{2013}]{Lewis:2013hha}
Lewis A.,  2013, \mn@doi [Physical Review D] {10.1103/PhysRevD.87.103529}, 87, 103529

\bibitem[\protect\citeauthoryear{Lewis}{Lewis}{2019}]{Lewis:2019xzd}
Lewis A.,  2019 (\mn@eprint {arXiv} {1910.13970})

\bibitem[\protect\citeauthoryear{Lewis \& Bridle}{Lewis \& Bridle}{2002}]{Lewis:2002ah}
Lewis A.,  Bridle S.,  2002, \mn@doi [Physical Review D] {10.1103/PhysRevD.66.103511}, 66, 103511

\bibitem[\protect\citeauthoryear{Lewis \& Ibata}{Lewis \& Ibata}{2002}]{Lewis:2002jt}
Lewis G.~F.,  Ibata R.~A.,  2002, \mn@doi [Monthly Notices of the Royal Astronomical Society] {10.1046/j.1365-8711.2002.05797.x}, 337, 26

\bibitem[\protect\citeauthoryear{Lewis, Challinor  \& Lasenby}{Lewis et~al.}{2000}]{Lewis:1999bs}
Lewis A.,  Challinor A.,   Lasenby A.,  2000, \mn@doi [The Astrophysical Journal] {10.1086/309179}, 538, 473

\bibitem[\protect\citeauthoryear{Linder}{Linder}{2003}]{Linder:2002et}
Linder E.~V.,  2003, \mn@doi [Physical Review Letters] {10.1103/PhysRevLett.90.091301}, 90, 091301

\bibitem[\protect\citeauthoryear{Motta, Garc\'\i{}a-Aspeitia, Hern\'andez-Almada, Maga\~na  \& Verdugo}{Motta et~al.}{2021}]{Motta:2021hvl}
Motta V.,  Garc\'\i{}a-Aspeitia M.~A.,  Hern\'andez-Almada A.,  Maga\~na J.,   Verdugo T.,  2021, \mn@doi [Universe] {10.3390/universe7060163}, 7, 163

\bibitem[\protect\citeauthoryear{{Neal}}{{Neal}}{2005}]{Neal:2005}
{Neal} R.~M.,  2005 (\mn@eprint {arXiv} {math/0502099})

\bibitem[\protect\citeauthoryear{{Perlmutter} et~al.}{{Perlmutter} et~al.}{1999}]{Perlmutter1999}
{Perlmutter} S.,  et~al., 1999, \mn@doi [The Astrophysical Journal] {10.1086/307221}, \href {http://adsabs.harvard.edu/abs/1999ApJ...517..565P} {517, 565}

\bibitem[\protect\citeauthoryear{Refsdal}{Refsdal}{1964}]{Refsdal1964b}
Refsdal S.,  1964, \mn@doi [Monthly Notices of the Royal Astronomical Society] {10.1093/mnras/128.4.307}, 128, 307

\bibitem[\protect\citeauthoryear{{Riess} et~al.}{{Riess} et~al.}{1998}]{Riess1998}
{Riess} A.~G.,  et~al., 1998, \mn@doi [The Astronomical Journal] {10.1086/300499}, \href {http://adsabs.harvard.edu/abs/1998AJ....116.1009R} {116, 1009}

\bibitem[\protect\citeauthoryear{Ross, Samushia, Howlett, Percival, Burden  \& Manera}{Ross et~al.}{2015}]{Ross:2014qpa}
Ross A.~J.,  Samushia L.,  Howlett C.,  Percival W.~J.,  Burden A.,   Manera M.,  2015, \mn@doi [Monthly Notices of the Royal Astronomical Society] {10.1093/mnras/stv154}, 449, 835

\bibitem[\protect\citeauthoryear{Rubakov}{Rubakov}{2014}]{Rubakov:2014jja}
Rubakov V.~A.,  2014, \mn@doi [Physics-Uspekhi] {10.3367/UFNe.0184.201402b.0137}, 57, 128

\bibitem[\protect\citeauthoryear{Schneider \& Sluse}{Schneider \& Sluse}{2013}]{Schneider:2013sxa}
Schneider P.,  Sluse D.,  2013, \mn@doi [Astronomy \& Astrophysics] {10.1051/0004-6361/201321882}, 559, A37

\bibitem[\protect\citeauthoryear{Schneider \& Sluse}{Schneider \& Sluse}{2014}]{Schneider:2013wga}
Schneider P.,  Sluse D.,  2014, \mn@doi [Astronomy \& Astrophysics] {10.1051/0004-6361/201322106}, 564, A103

\bibitem[\protect\citeauthoryear{Scolnic et~al.}{Scolnic et~al.}{2018}]{Scolnic:2017caz}
Scolnic D.~M.,  et~al., 2018, \mn@doi [The Astrophysical Journal] {10.3847/1538-4357/aab9bb}, 859, 101

\bibitem[\protect\citeauthoryear{Shiralilou, Martinelli, Papadomanolakis, Peirone, Renzi  \& Silvestri}{Shiralilou et~al.}{2020}]{Shiralilou:2019div}
Shiralilou B.,  Martinelli M.,  Papadomanolakis G.,  Peirone S.,  Renzi F.,   Silvestri A.,  2020, \mn@doi [Journal of Cosmology and Astroparticle Physics] {10.1088/1475-7516/2020/04/057}, 04, 057

\bibitem[\protect\citeauthoryear{Sonnenfeld}{Sonnenfeld}{2018}]{Sonnenfeld:2017dca}
Sonnenfeld A.,  2018, \mn@doi [Monthly Notices of the Royal Astronomical Society] {10.1093/mnras/stx3105}, 474, 4648

\bibitem[\protect\citeauthoryear{Torrado \& Lewis}{Torrado \& Lewis}{2021}]{Torrado:2020dgo}
Torrado J.,  Lewis A.,  2021, \mn@doi [Journal of Cosmology and Astroparticle Physics] {10.1088/1475-7516/2021/05/057}, 05, 057

\bibitem[\protect\citeauthoryear{{Walsh}, {Carswell}  \& {Weymann}}{{Walsh} et~al.}{1979}]{Walsh1979}
{Walsh} D.,  {Carswell} R.~F.,   {Weymann} R.~J.,  1979, \mn@doi [Nature] {10.1038/279381a0}, \href {https://ui.adsabs.harvard.edu/abs/1979Natur.279..381W} {279, 381}

\bibitem[\protect\citeauthoryear{Wang, Zhang, He, Zhang  \& Zhang}{Wang et~al.}{2022}]{Wang:2021kxc}
Wang L.-F.,  Zhang J.-H.,  He D.-Z.,  Zhang J.-F.,   Zhang X.,  2022, \mn@doi [Monthly Notices of the Royal Astronomical Society] {10.1093/mnras/stac1468}, 514, 1433

\bibitem[\protect\citeauthoryear{Wong et~al.}{Wong et~al.}{2020}]{Wong:2019kwg}
Wong K.~C.,  et~al., 2020, \mn@doi [Monthly Notices of the Royal Astronomical Society] {10.1093/mnras/stz3094}, 498

\makeatother
\end{thebibliography}

\bsp
\label{lastpage}
\end{document}